\documentclass[aps,prl,showpacs,nofootinbib,nobalancelastpage,twocolumn]{revtex4}
\usepackage{amssymb}
\usepackage{amsfonts}
\usepackage{amsmath,color}
\usepackage{dsfont}
\usepackage{epsfig}
\usepackage{graphicx,epsf,epsfig}
\usepackage{bbm}
\usepackage{subfigure}
\usepackage{multirow}
\usepackage{setspace}
\usepackage{verbatim}

\begin{document}

\title{Non-Hermitian neutrino oscillations in matter with PT symmetric Hamiltonians}

\author{Tommy Ohlsson}
\email{tohlsson@kth.se}
\affiliation{Department of Theoretical Physics, School of Engineering Sciences, KTH Royal Institute
of Technology, AlbaNova University Center -- Roslagstullsbacken 21, 106~91 Stockholm, Sweden}

\begin{abstract}
We introduce and develop a novel approach to extend the ordinary two-flavor neutrino oscillation formalism in matter using a non-Hermitian PT symmetric effective Hamiltonian. The condition of PT symmetry is weaker and less mathematical than that of hermicity, but more physical, and such an extension of the formalism can give rise to sub-leading effects in neutrino flavor transitions similar to the effects by so-called non-standard neutrino interactions. We derive the necessary conditions for the spectrum of the effective Hamiltonian to be real as well as the mappings between the fundamental and effective parameters. We find that the real spectrum of the effective Hamiltonian will depend on all new fundamental parameters introduced in the non-Hermitian PT symmetric extension of the usual neutrino oscillation formalism and that either i) the spectrum is exact and the effective leptonic mixing must always be maximal or ii) the spectrum is approximate and all new fundamental parameters must be small.
\end{abstract}

\pacs{14.60.Pq, 11.30.Er, 13.15.+g}

\maketitle

In 1998, the Super-Kamiokande collaboration \cite{Fukuda:1998mi} reported results that gave a clear indication of neutrino oscillations ---a quantum mechanical effect over distances of thousands of kilometers, and thus the first solid evidence for physics beyond the Standard Model of particle physics. Now, it has been firmly established that neutrino oscillations is the best and \emph{leading} description of neutrino flavor transitions and Takaaki Kajita and Arthur B.~McDonald have been awarded the Nobel Prize in Physics 2015 ``for the discovery of neutrino oscillations \ldots''. However, other mechanisms could be responsible for such transitions on a \emph{sub-leading} level. Therefore, other ``new physics'' effects have been proposed, \emph{e.g.}~so-called non-standard neutrino interactions (NSIs). See, \emph{e.g.}, ref.~\cite{Ohlsson:2012kf} and references therein.

Nevertheless, in the same year as the Super-Kamiokande collaboration reported its results, Bender and Boettcher  \cite{Bender:1998ke} presented (based on a conjecture by Bessis) PT symmetric quantum mechanics, which is described by non-Hermitian Hamiltonians. In ordinary quantum mechanics, a Hamiltonian operator $H$ should be Hermitian ($H^\dagger = H$) in order to have real (and measurable) energy eigenvalues. However, using PT symmetric quantum mechanics, the requirement of hermicity can be replaced by a weaker, but more physical, requirement of space-time reflection symmetry (so-called PT symmetry, \emph{i.e.}~$[{\rm PT},H] = 0$) without losing any important physical aspects of ordinary quantum mechanics \cite{Bender:2005tb}. Note that P is parity (space reflection) and T is time reversal. Now, PT is an antiunitary operator \cite{Porter:1965}. The definition for any such operator $A$ is that $\langle A \phi | A \psi \rangle = \langle \phi | \psi \rangle^*$, where $| \psi \rangle$ and $| \phi \rangle$ are arbitrary states. Any antiunitary operator can be written in the form $A = {\cal U}K$, where ${\cal U}$ is a unitary operator (\emph{e.g.}, P) and $K$ is a complex conjugation (\emph{e.g.}, T in position representation). Consider Hamiltonians with antiunitary symmetry, \emph{i.e.}~$[A,H] = 0$, where $A^{2k} = 1$ ($k$ odd). Note that PT corresponds to $k = 1$. It has be shown that for any such $A$, it is possible to construct a basis in which the matrix elements of $H$ are real, which means that the characteristic equation giving the eigenvalues of $H$ is real \cite{Bender:2002yp}. This result is a generalization of applications in quantum chaology \cite{Robnik:1986}, which, in turn, was a generalization of arguments from nuclear physics \cite{Porter:1965}.

In refs.~\cite{JonesSmith:2009wy,JonesSmith:2010,White:2010}, it has been shown that two Dirac fermions coupled by a non-Hermitian PT symmetric mass matrix describes a single 8-dimensional relativistic particle of a fundamentally new type that can appear as two massless particles, despite the mass matrix being non-zero. It is then evident that a non-Hermitian PT symmetric mass matrix might lead to a different neutrino oscillation phenomenology, since such a mass matrix could describe neutrinos that oscillate between two flavors (given by the two massless particles) but propagate masslessly. Now, evidence of neutrino oscillations would normally mean that neutrinos are massive. However, in PT symmetric quantum mechanics, the 8-dimensional solution suggests that this conclusion might not need to be drawn. Later, in ref.~\cite{Berryman:2014yoa}, it has been discussed that interactions between ordinary light neutrino states and new very light many-particle states could lead to new physics described by non-Hermitian Hamiltonians. The idea has been developed for neutral kaons and can be readily adapted to neutrinos, which could interact with and decay into such generic light states. For neutrinos, there are no constraints from CPT invariance, whereas for neutral kaons, this is not the case. Recently, in ref.~\cite{Alexandre:2015kra},  a non-Hermitian Yukawa theory has been studied that could give an explanation for the smallness of the masses of light neutrinos. Finally, connecting PT symmetric quantum mechanics to neutrino oscillations, Bender {\it et al.}~have experimentally studied a simple mechanical system consisting of two coupled pendula, which is the classical analog of the phenomenon of neutrino oscillations. They have observed a phase transition that separates the unbroken and broken phases of the PT symmetry of the quantum-mechanical Hamiltonian \cite{Bender:2013}. Therefore, this simple experiment and its result provide an intuitive motivation to study non-Hermitian PT symmetric Hamiltonians for neutrino flavor transitions in matter.

In this Letter, we will introduce and investigate two-flavor neutrino oscillations in matter (of constant density) based on non-Hermitian, but PT symmetric, Hamiltonians that could give rise to sub-leading effects in neutrino flavor transitions. Based on the physical motivation of earlier results \cite{JonesSmith:2009wy,JonesSmith:2010,White:2010,Berryman:2014yoa,Alexandre:2015kra,Bender:2013}, and especially the result of ref.~\cite{Bender:2013}, such an extension of ordinary neutrino oscillations will therefore open up the possibility for new physics beyond the Standard Model. In general, non-Hermitian extensions relax assumptions about constraints made on Hermitian models (such as the requirement of hermicity) rather than adding new dynamical degrees of freedom. Thus, they constitute a different paradigm of new physics compared to traditional model building.

Assuming two lepton flavors (for generality, they will be denoted $\alpha$ and $\beta$), the time evolution of the neutrino vector of state $\nu = \left( \begin{matrix} \nu_\alpha & \nu_\beta \end{matrix} \right)^{\rm T}$ describing neutrino oscillations is given by a Schr{\"o}dinger-like equation with a time-independent Hermitian Hamiltonian $H_{\rm osc}$, namely
\begin{align}
{\rm i} \frac{{\rm d} \nu(t)}{{\rm d} t} &= \frac{1}{2E} \left[ O \left( \begin{matrix} m_1^2 & 0 \\ 0 & m_2^2 \end{matrix} \right) O^{\rm T} + \left( \begin{matrix} A & 0 \\ 0 & 0 \end{matrix} \right) \right] \nu(t) \nonumber\\
&\equiv H_{\rm osc} \nu(t) \,,
\end{align}
where $E$ is the neutrino energy and $A = 2 \sqrt{2} E G_F N_e$ is the effective matter potential induced by ordinary charged-current weak interactions with electrons \cite{Wolfenstein:1977ue,Mikheev:1986gs}. Here, $m_1$ and $m_2$ are the definite masses of the neutrino mass eigenstates, $\nu_1$ and $\nu_2$, respectively, that are related to the weak interaction eigenstates, $\nu_\alpha$ and $\nu_\beta$, through the leptonic mixing matrix $O$ such that $\left( \begin{matrix} \nu_\alpha & \nu_\beta \end{matrix} \right)^{\rm T} = O \left( \begin{matrix} \nu_1 & \nu_2 \end{matrix} \right)^{\rm T}$, $G_F$ is the Fermi coupling constant, and $N_e$ is the electron density of matter along the neutrino trajectory. Furthermore, the two-flavor leptonic mixing matrix (parametrized by the leptonic mixing angle $\theta$, \emph{i.e.}~one real parameter) can be written as
\begin{equation}
O = \left( \begin{matrix} c & s \\ -s & c \end{matrix} \right) \in SO(2)\,,
\end{equation}
where $c \equiv \cos \theta$ and $s \equiv \sin \theta$. Using the formalism of neutrino oscillations, the quantum mechanical transition probability amplitudes are given as overlaps of different neutrino states, and eventually, neutrino oscillation probabilities are defined as squared absolute values of the amplitudes. Thus, neutrino flavor transitions occur during the evolution of neutrinos. In vacuum (\emph{i.e.}~assuming $N_e = 0$ (or $A = 0$)), using conservation of probability (\emph{i.e.}~unitarity), the well-known two-flavor neutrino oscillations probability formulas are given by (see, \emph{e.g.}, ref.~\cite{Bilenky:1987ty})
\begin{align}
&P(\nu_\alpha \to \nu_\beta; L) = \sin^2(2\theta) \sin^2 \left( \frac{\Delta m^2 L}{4E} \right) \,, \label{eq:Pab_vac}\\
&P(\nu_\alpha \to \nu_\alpha; L) = 1 - P(\nu_\alpha \to \nu_\beta; L) \nonumber\\
&= 1 - P(\nu_\beta \to \nu_\alpha; L) = P(\nu_\beta \to \nu_\beta; L) \,, \label{eq:Paa_vac}
\end{align}
where $L$ is the (propagation) path length of the neutrinos and $\Delta m^2 \equiv m_2^2 - m_1^2$ is the mass-squared difference between the masses of the two neutrino mass eigenstates. Note that $\theta$ corresponds to the amplitude of the oscillations, whereas $\Delta m^2$ corresponds to the frequency. Indeed, in matter of constant density (\emph{i.e.}~assuming $N_e = {\rm const.} \neq 0$) as well as in the case of NSIs, the probability formulas are obtained by replacing the vacuum parameters with effective matter or NSI para\-meters in eqs.~(\ref{eq:Pab_vac}) and (\ref{eq:Paa_vac}). (Note that if $N_e = {\rm const.}$, $A \propto E$, and hence, $A$ is only constant for a fixed $E$.) Thus, we have the transition probability
\begin{equation}
P(\nu_\alpha \to \nu_\beta; L) = \sin^2(2\theta') \sin^2 \left( \frac{{\Delta m'}^2 L}{4E} \right) \,, \label{eq:Pab_mat}
\end{equation}
where $\theta'$ is the effective mixing angle and ${\Delta m'}^2$ is the effective mass-squared difference. Therefore, there are non-trivial mappings between the vacuum parameters $\theta$ and $\Delta m^2$ and the corresponding effective (matter or NSI) parameters $\theta'$ and ${\Delta m'}^2$. The key point will be the diagonalization of a given effective Hamiltonian $H_{\rm eff}$ and obtaining the explicit relations of the effective parameters in terms of the fundamental (vacuum) parameters \cite{Meloni:2009ia}.

In the case of matter of constant density, the mappings are given by
\begin{align}
\sin^2(2\theta') &= \frac{\sin^2(2\theta)}{\sin^2(2\theta) + \left( \cos(2\theta) - A/\Delta m^2 \right)^2} \,, \label{eq:theta'}\\
{\Delta m'}^2 &= \sqrt{(\Delta m^2)^2 \sin^2(2\theta) + \left( \Delta m^2 \cos(2\theta) - A \right)^2} \,. \label{eq:m'}
\end{align}
Note that the effective matter mixing angle in eq.~(\ref{eq:theta'}) obtains its maximal value $\sin^2(2\theta') = 1$ when the condition
\begin{equation}
A = \Delta m^2 \cos(2\theta)
\label{eq:MSW}
\end{equation}
is fulfilled. This condition is the famous Mikheyev-Smirnov-Wolfenstein (MSW) resonance condition \cite{Mikheev:1986gs,Mikheev:1986wj}. If the MSW condition is satisfied, then the effective matter mixing angle $\theta'$ is maximal ($\theta' = 45^\circ$) independently of the value of the fundamental vacuum mixing angle $\theta$. Thus, the transition probability in eq.~(\ref{eq:Pab_mat}) can be very large even if $\theta$ is very small. In the vacuum limit, \emph{i.e.}~the limit of vanishing $A$, the effective parameters are replaced by the fundamental vacuum parameters as well as eqs.~(\ref{eq:Pab_vac}) and (\ref{eq:Pab_mat}) are identical to each other and the mappings~(\ref{eq:theta'}) and (\ref{eq:m'}) are reduced to $\theta' = \theta$ and ${\Delta m'}^2 = \left| \Delta m^2 \right|$, respectively.

In order to study the features of adding a non-Hermitian, but PT symmetric, Hamiltonian to the ordinary formalism of neutrino oscillations in matter, we consider for illustration the following general complex Hamiltonian \cite{Bender:2005tb,Alexandre:2015kra}:
\begin{equation}
H' = \frac{A}{2E} \left( \begin{matrix} \rho {\rm e}^{{\rm i} \varphi} & \sigma \\ \sigma & \rho {\rm e}^{-{\rm i} \varphi} \end{matrix} \right) \,, \label{eq:H'}
\end{equation}
where $\rho$, $\sigma$, and $\varphi$ are three real parameters. The physical meaning of $\rho$, $\sigma$, and $\varphi$ is similar to other phenomenologi\-cal parameters for ``new physics'' (\emph{e.g.}, NSI parameters). The Hamiltonian $H'$ is not Hermitian, but it is PT symmetric. However, note that if $\varphi = 0,\pi$, then $H'$ becomes Hermitian. Using the characteristic equation of this Hamiltonian, \emph{i.e.}~$\det (2E H' - \lambda {\mathds 1}_2) = 0$, we find the two eigenvalues $\lambda = A \rho \cos \varphi \pm A \sqrt{\sigma^2 - \rho^2 \sin^2 \varphi}$. Thus, there are two parametric regions. If $\sigma^2 < \rho^2 \sin^2 \varphi$, the two eigenvalues form a complex conjugate pair and the PT symmetry is \emph{broken}, whereas if $\sigma^2 > \rho^2 \sin^2 \varphi$, the two eigenvalues are real and the PT symmetry is \emph{unbroken}. At the point between the two regions, \emph{i.e.}~$\sigma^2 = \rho^2 \sin^2 \varphi$, there is only one eigenvalue and it is called an \emph{exceptional point}. The interesting region to consider is the unbroken region with conservation of PT symmetry and two real eigenvalues. Note that complex Hamiltonians may be either Hermitian or PT symmetric, but not both, whereas real symmetric Hamiltonians may be both Hermitian and PT symmetric \cite{Bender:2005tb}. Another way of introducing a complex non-Hermitian Hamiltonian in a different context has been performed for high-energy two-flavor neutrino oscillations in matter with absorption leading to complex indices  of refraction \cite{Naumov:2001ci}.

Now, constructing the effective Hamiltonian $H_{\rm eff}$ by adding the non-Hermitian PT symmetric Hamiltonian $H'$ to the ordinary neutrino oscillation Hamiltonian in matter $H_{\rm osc}$ and following refs.~\cite{Bilenky:1987ty,Kim:1994dy}, we obtain the total Hamiltonian, which is complex symmetric, as
\begin{align}
H_{\rm eff} &= H_{\rm osc} + H' \nonumber\\
&= \frac{1}{2E} \Bigg[  O \left( \begin{matrix} m_1^2 & 0 \\ 0 & m_2^2 \end{matrix} \right) O^{\rm T} + \left( \begin{matrix} A & 0 \\ 0 & 0 \end{matrix} \right) \nonumber\\
&+ A \left( \begin{matrix} \rho {\rm e}^{{\rm i} \varphi} & \sigma \\ \sigma & \rho {\rm e}^{-{\rm i} \varphi} \end{matrix} \right) \Bigg] \nonumber\\
&= \frac{1}{2E} \tilde{U} \left( \begin{matrix} \tilde{m}_1^2 & 0 \\ 0 & \tilde{m}_2^2 \end{matrix} \right) \tilde{U}^{\rm T} \,, \label{eq:Heff}
\end{align}
where $\tilde{U}$ is the effective leptonic mixing matrix and $\tilde{m}_1$ and $\tilde{m}_2$ are the two effective neutrino masses, which are assumed to be non-degenerate, otherwise there would not be any oscillations. Since $H_{\rm eff}$ is a $2 \times 2$ complex symmetric matrix, $\tilde{U}$ must be a $2 \times 2$ unitary matrix such that
\begin{equation}
\tilde{U} = \left( \begin{matrix} \tilde{c} {\rm e}^{{\rm i} \tilde{\varphi}_1} & \tilde{s} {\rm e}^{{\rm i} \tilde{\varphi}_2} \\ -\tilde{s} {\rm e}^{-{\rm i} \tilde{\varphi}_2} & \tilde{c} {\rm e}^{-{\rm i} \tilde{\varphi}_1} \end{matrix} \right) \in SU(2) \,,
\end{equation}
where $\tilde{c} \equiv \cos \tilde{\theta}$, $\tilde{s} \equiv \sin \tilde{\theta}$, $\tilde{\theta}$ being the effective mixing angle, and $\tilde{\varphi}_1$ and $\tilde{\varphi}_2$ are two phases, \emph{i.e.}~in total three real parameters. Note that the total Hamiltonian can be expressed in three different useful and relevant bases of the same Hilbert space, namely the flavor, mass, and effective bases. Thus, we have $H_{\rm eff} = \tilde{U} H_e \tilde{U}^{\rm T} = O H_m O^{-1}$, where $H_e$ and $H_m$ are the total Hamiltonian in effective basis and mass basis, respectively. In the case of two lepton flavors, $O$ is orthogonal, and therefore, $O^{-1} = O^{\rm T}$. Thus, diagonalizing the effective total Hamiltonian $H_{\rm eff}$ in eq.~(\ref{eq:Heff}), we have the following six relations:
\begin{widetext}
\begin{align}
c^2 m_1^2 + s^2 m_2^2 + A + A \rho \cos \varphi &= \tilde{c}^2 \cos(2\tilde{\varphi}_1) \tilde{m}_1^2 + \tilde{s}^2 \cos(2\tilde{\varphi}_2) \tilde{m}_2^2 \,, \label{eq:rel1}\\
A \rho \sin \varphi &= \tilde{c}^2 \sin(2\tilde{\varphi}_1) \tilde{m}_1^2 + \tilde{s}^2 \sin(2\tilde{\varphi}_2) \tilde{m}_2^2 \,, \\
sc \Delta m^2 + A \sigma &= \tilde{s} \tilde{c} \cos(\tilde{\varphi}_1 - \tilde{\varphi}_2) \Delta \tilde{m}^2 \,, \\
0 &= \tilde{s} \tilde{c} \sin(\tilde{\varphi}_1 - \tilde{\varphi}_2) \Delta \tilde{m}^2 \,, \label{eq:rel4} \\
s^2 m_1^2 + c^2 m_2^2 + A \rho \cos \varphi &= \tilde{s}^2 \cos(2\tilde{\varphi}_2) \tilde{m}_1^2 + \tilde{c}^2 \cos(2\tilde{\varphi}_1) \tilde{m}_2^2 \,, \label{eq:rel5}\\
A \rho \sin \varphi &= \tilde{s}^2 \sin(2\tilde{\varphi}_2) \tilde{m}_1^2 + \tilde{c}^2 \sin(2\tilde{\varphi}_1) \tilde{m}_2^2 \,, \label{eq:rel6}
\end{align}
\end{widetext}
where $\Delta \tilde{m}^2 \equiv \tilde{m}_2^2 - \tilde{m}_1^2$. Note that eq.~(\ref{eq:rel4}) immediately implies that $\tilde{\varphi}_1 = \tilde{\varphi}_2$. Solving eqs.~(\ref{eq:rel1})--(\ref{eq:rel6}) for the effective amplitude and frequency, we find the possible mappings
\begin{align}
\sin^2 (2\tilde{\theta}) &= \frac{\left(\Delta m^2 \sin(2\theta) + 2 A \sigma \right)^2}{\left(\Delta m^2 \sin(2\theta) + 2 A \sigma \right)^2 + \left(\Delta m^2 \cos(2\theta) - A\right)^2} \,, \label{eq:amp}\\
\Delta \tilde{m}^2 &= \sqrt{\left(\Delta m^2 \sin(2\theta) + 2 A \sigma \right)^2 + \left(\Delta m^2 \cos(2\theta) - A\right)^2} \,. \label{eq:freq}
\end{align}
In addition, we obtain two auxiliary mappings
\begin{align}
\tan(2\tilde{\varphi}) &= \frac{2 A \rho \sin \varphi}{m_1^2 + m_2^2 + A + 2 A \rho \cos \varphi} \,, \label{eq:auxphase}\\
\tilde{m}_1^2 + \tilde{m}_2^2 &= \sqrt{\left( m_1^2 + m_2^2 + A + 2 A \rho \cos \varphi \right)^2 + 4 A^2 \rho^2 \sin^2 \varphi} \,, \label{eq:auxsum}
\end{align}
where $\tilde{\varphi} \equiv \tilde{\varphi}_1 = \tilde{\varphi}_2$. In the vacuum limit (\emph{i.e.}~$A \to 0$), for eqs.~(\ref{eq:amp})--(\ref{eq:auxsum}), we have $\tilde{\theta} = \theta$, $\Delta \tilde{m}^2 = \left| \Delta m^2 \right|$, $\tilde{\varphi} = 0$, and $\tilde{m}_1^2 + \tilde{m}_2^2 = \left| m_1^2 + m_2^2 \right|$. In this case, note that the effective phase $\tilde{\varphi}$ naturally becomes equal to zero, since it does not have a correspondence in ordinary vacuum neutrino oscillations. 

Does the effective total Hamiltonian $H_{\rm eff}$ have a region of unbroken PT symmetry? In order to answer this question, we need to consider the characteristic equation of this Hamiltonian, \emph{i.e.}~$\det (2E H_{\rm eff} - \lambda {\mathds 1}_2) = 0$, which leads to a quadratic equation
\begin{align}
&\lambda^2 - (m_1^2 + m_2^2 + A + 2 A \rho \cos \varphi) \lambda \nonumber\\
&+ (c^2 m_1^2 + s^2 m_2^2 + A)(s^2 m_1^2 + c^2 m_2^2) \nonumber\\
&+ A^2 \rho^2 - (sc \Delta m^2 + A \sigma)^2 \nonumber\\
&+ (s^2 m_1^2 + c^2 m_2^2) A \rho {\rm e}^{{\rm i} \varphi} \nonumber\\
&+ (c^2 m_1^2 + s^2 m_2^2 + A) A \rho {\rm e}^{-{\rm i} \varphi} = 0 \,, \label{eq:chareq}
\end{align}
where the last two terms on the left-hand side of this equation contain imaginary parts. The condition for the equation to have real roots is that the imaginary parts equal zero, see, \emph{e.g.}, ref.~\cite{Hardy:1921}, pp.~91--92. It turns out that the imaginary parts can be reduced to $\Im = A \rho \sin \varphi (\Delta m^2 \cos(2\theta) - A)$, which means that $\Im = 0$ if the MSW condition in eq.~(\ref{eq:MSW}) is fulfilled. Of course, $\Im = 0$ when either $\rho = 0$ or $\varphi = 0,\pi$, but these are trivial cases and therefore not interesting. (Later, we will investigate another non-trivial, but approximate, case.) Thus, the MSW condition is a necessary constraint for having two potential real \emph{exact} eigenvalues of $H_{\rm eff}$.

Inserting the MSW condtion~(\ref{eq:MSW}) into eq.~(\ref{eq:chareq}), we obtain
\begin{align}
\lambda &= s^2 m_1^2 + c^2 m_2^2 + A \rho \cos \varphi \nonumber \\
&\pm \sqrt{(sc \Delta m^2 + A \sigma)^2 - A^2 \rho^2 \sin^2 \varphi} \,. \label{eq:lambda}
\end{align}
In order for eq.~(\ref{eq:lambda}) to have two real eigenvalues, the quadratic form $(sc \Delta m^2 + A \sigma)^2 - A^2 \rho^2 \sin^2 \varphi$ must be positive definite, which consecutively means that the following condition needs to be fulfilled:
\begin{equation}
\left| \Delta m^2 \sin(2\theta) + 2 A \sigma \right| > \left| 2 A \rho \sin \varphi \right| \,. \label{eq:PT}
\end{equation}
Thus, $H_{\rm eff}$ has a region of unbroken PT symmetry if the two conditions in eqs.~(\ref{eq:MSW}) and (\ref{eq:PT}) are satisfied. In this case, using eqs.~(\ref{eq:amp}) and (\ref{eq:freq}), the mappings for the effective mixing angle and the effective mass-squared difference, respectively, are reduced to
\begin{align}
\sin^2(2\tilde{\theta}) &= 1 \,, \label{eq:ampPT}\\
\Delta \tilde{m}^2 &= \left| \Delta m^2 \sin(2\theta) + 2 A \sigma \right| > \left| 2 A \rho \sin \varphi \right| \,. \label{eq:freqPT}
\end{align}
We observe that the value of the effective mixing angle is always maximal ($\tilde{\theta} = 45^\circ$) independently of the values of the three parameters $\rho$, $\sigma$, and $\varphi$ describing $H'$. The reason is that the MSW condition has to be satisfied in eq.~(\ref{eq:amp}). Furthermore, we note that the effective mass-squared difference is always positive and only linearly dependent on the parameter $\sigma$.

The effects of a non-Hermitian PT symmetric Hamiltonian on neutrino flavor transitions must be sub-leading, and therefore, the parameters describing $H'$ in eq.~(\ref{eq:H'}) have to be small. Thus, we series expand the two auxiliary mappings~(\ref{eq:auxphase}) and (\ref{eq:auxsum}) up to second order in the small parameter $\rho$ (note that the parameter $\varphi$ is a phase):
\begin{align}
\tilde{\varphi} &= \frac{A \sin \varphi}{m_1^2 + m_2^2 + A} \rho + {\cal O}(\rho^2) \,, \label{eq:seriesphase} \\
\tilde{m}_1^2 + \tilde{m}_2^2 &= \left| m_1^2 + m_2^2 + A \right| \nonumber\\
&+ 2 \, {\rm sgn}(m_1^2 + m_2^2 + A) A \rho \cos \varphi + {\cal O}(\rho^2) \,, \label{eq:seriessum}
\end{align}
which are both directly independent of the parameter $\sigma$. Indeed, we observe that the effective phase in eq.~(\ref{eq:seriesphase}) has a sinusoidal dependence on $\varphi$ for small $\rho$. In addition, using eqs.~(\ref{eq:freqPT}) and (\ref{eq:seriessum}), we obtain the series expansions of the two real eigenvalues of $H_{\rm eff}$ as
\begin{align}
\tilde{m}_1^2 &= \frac{1}{2} (\left| m_1^2 + m_2^2 + A \right| - \left| \Delta m^2 \sin(2\theta) + 2 A \sigma \right|) \nonumber\\
&+ {\rm sgn}(m_1^2 + m_2^2 + A) A \rho \cos \varphi + {\cal O}(\rho^2) \,, \label{eq:seriesm12}\\
\tilde{m}_2^2 &= \frac{1}{2} (\left| m_1^2 + m_2^2 + A \right| + \left| \Delta m^2 \sin(2\theta) + 2 A \sigma \right|) \nonumber\\
&+ {\rm sgn}(m_1^2 + m_2^2 + A) A \rho \cos \varphi + {\cal O}(\rho^2) \,. \label{eq:seriesm22}
\end{align}

Now, combining the MSW condition~(\ref{eq:MSW}) and the PT inequality~(\ref{eq:PT}), we obtain a single inequality among the fundamental parameters in order to have a non-Hermitian PT symmetric effective total Hamiltonian with a real \emph{exact} spectrum, namely
\begin{equation}
A^2 \left[ 4 (\rho \sin \varphi - \sigma)^2 + 1 \right] < (\Delta m^2)^2 \,, \label{eq:singlecond}
\end{equation}
which holds if both $\Delta m^2 \sin(2\theta) + 2 A \sigma$ and $A \rho \sin \varphi$ are positive or negative. In the case in which one of these two expressions is positive and the other one is negative or vice versa, the corresponding inequality is
\begin{equation}
A^2 \left[ 4 (\rho \sin \varphi + \sigma)^2 + 1 \right] < (\Delta m^2)^2 \,. \label{eq:singlecond2}
\end{equation}
Since $\Delta m^2$ (or equivalently, $\theta$ via $\cos(2\theta) = A/\Delta m^2 = 2\sqrt{2} G_F N_e E_{\rm res}/\Delta m^2$, where $E_{\rm res}$ is the resonance energy) is the considered two-flavor neutrino mass-squared difference (given by Nature) and $A$ is the effective matter potential for the chosen experimental setup (\emph{e.g.}, a specific neutrino oscillation experiment), the model parameters $\rho$, $\sigma$, and $\varphi$ have to be restricted such that eq.~(\ref{eq:singlecond}) (or eq.~(\ref{eq:singlecond2})) is fulfilled. Thus, the important inequality~(\ref{eq:singlecond}) sets the bound on the new fundamental parameters that still allows the effective total Hamiltonian~(\ref{eq:Heff}) to have an unbroken PT symmetry with real \emph{exact} eigenvalues.

For example, in the case of atmospheric neutrinos, we have the parameter values $\theta \simeq \theta_{13} = 8.5^\circ$, $\Delta m^2 \simeq \Delta m^2_{31} = 2.5 \cdot 10^{-3} \; {\rm eV}^2$, and $E_{\rm res}^{\rm atm} \simeq 11 \; {\rm GeV}$, whereas in the case of solar neutrinos, we have $\theta \simeq \theta_{12} = 33^\circ$, $\Delta m^2 \simeq \Delta m^2_{21} = 7.5 \cdot 10^{-5} \; {\rm eV}^2$, and $E_{\rm res}^\odot \simeq 0.13 \; {\rm GeV}$ \cite{Akhmedov:2004ny,Gonzalez-Garcia:2014bfa}. Thus, inserting the two sets of parameter values into eq.~(\ref{eq:singlecond}), we find the following estimates of the upper bounds on the model parameters: $\rho \sin \varphi - \sigma \lesssim 0.15$ (for atmospheric neutrinos) and $\rho \sin \varphi - \sigma \lesssim 1.1$ (for solar neutrinos).

\begin{figure}
\begin{center}
\includegraphics[width=0.5\textwidth]{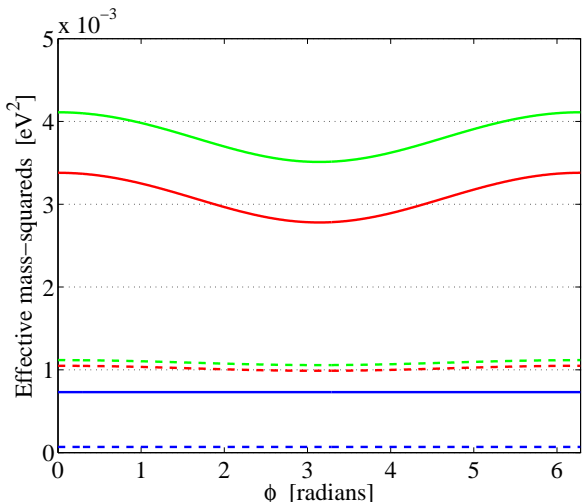}
\end{center}
\vspace{-0.75cm}
\caption{The spectrum of the effective mass-squareds $\tilde{m}_1^2$ and $\tilde{m}_2^2$ as functions of $\varphi \in [0,2\pi)$ with $\sigma = 0$ and $m_1^2 = 10^{-3} \; {\rm eV}^2$. The solid curves show the case of atmospheric neutrinos ($\rho = 0.1$), whereas the dashed curves show the case of solar neutrinos ($\rho = 1$). A red curve depicts $\tilde{m}_1^2$, whereas a green curve depicts $\tilde{m}_2^2$. In addition, a blue curve displays $\Delta \tilde{m}^2 = \tilde{m}_2^2 - \tilde{m}_1^2$. The fundamental neutrino oscillation parameters used are: $\Delta m_{31}^2 = 2.5 \cdot 10^{-3} \; {\rm eV}^2$, $\Delta m_{21}^2 = 7.5 \cdot 10^{-5} \; {\rm eV}^2$, $\theta_{13} = 8.5^\circ$, $\theta_{12} = 33^\circ$.}
\label{fig:m12_m22_phi}
\end{figure}

In fig.~\ref{fig:m12_m22_phi}, using eqs.~(\ref{eq:auxsum}) and (\ref{eq:freqPT}) as well as the fact that $m_1^2 + m_2^2 + A = 2 (m_1^2 + c^2 \Delta m^2)$, we present an illustration of the effective mass-squareds $\tilde{m}_1^2$ and $\tilde{m}_2^2$, which constitute the spectrum of $H_{\rm eff}$, for the cases of atmospheric and solar neutrino oscillations.
For completeness, we also display for the two cases the effective mass-squared difference given in eq.~(\ref{eq:freqPT}), which is independent of $\varphi$, and are therefore shown as straight lines in the figure.

We have shown that eqs.~(\ref{eq:lambda})--(\ref{eq:freqPT}) and (\ref{eq:seriesm12})--(\ref{eq:singlecond2}) are only valid at the MSW resonance described by the MSW condition~(\ref{eq:MSW}). However, note that eqs.~(\ref{eq:seriesphase}) and (\ref{eq:seriessum}) do not assume this condition and are therefore valid away from this resonance. What happens for an experimental setup that does not fulfill eq.~(\ref{eq:singlecond}) or (\ref{eq:singlecond2})? Which Hamiltonian should be used away from (or at least not close to) the MSW resonance? In fact, the same Hamiltonian, \emph{i.e.}~eq.~(\ref{eq:Heff}), should be used everywhere, but, as mentioned above, the effects of a non-Hermitian PT symmetric Hamiltonian on neutrino flavor transitions must be sub-leading, and therefore, the new fundamental parameters must be small. Assuming the three fundamental parameters $\rho$, $\sigma$, and $\varphi$ to be small (\emph{i.e.}~$\rho, \sigma, \varphi \ll 1$) and keeping only terms up to first order in perturbation theory, the imaginary parts in eq.~(\ref{eq:chareq}) also equal zero, since
\begin{align}
\Im &= A \rho \sin \varphi (\Delta m^2 \cos(2\theta) - A) \nonumber\\
&= A \rho \varphi (\Delta m^2 \cos(2\theta) - A) + {\cal O}(\rho \varphi^3) \simeq 0,
\end{align}
where the factor $\rho \varphi$ is second order in the small para\-meters. This is the other non-trivial case of obtaining a region of unbroken PT symmetry with two real eigenvalues of $H_{\rm eff}$. Therefore, in this case, eq.~(\ref{eq:chareq}) reduces to
\begin{align}
&\lambda^2 - (m_1^2 + m_2^2 + A + 2 A \rho) \lambda \nonumber\\
&+ (c^2 m_1^2 + s^2 m_2^2 + A)(s^2 m_1^2 + c^2 m_2^2) \nonumber\\
& - (sc \Delta m^2)(sc \Delta m^2 + 2 A \sigma) \simeq 0 \,. \label{eq:chareq2}
\end{align}
Note that this case is only approximate, whereas the other case described by eq.~(\ref{eq:singlecond}) or (\ref{eq:singlecond2}) is exact. However, instead of solving eq.~(\ref{eq:chareq2}) directly, we will perform the same approximations in eqs.~(\ref{eq:rel1})--(\ref{eq:auxsum}), which will lead to the desired results in a less tedious way. Now, using eq.~(\ref{eq:auxphase}) to observe that $\tilde{\varphi} \simeq 0$, since $\rho \sin \varphi = \rho \varphi + {\cal O}(\rho \varphi^3) \simeq 0$, and series expanding eqs.~(\ref{eq:rel1}) and (\ref{eq:rel5}) in the small parameters $\rho$ and $\varphi$, we obtain
\begin{align}
c^2 m_1^2 + s^2 m_2^2 + A + A \rho &\simeq \tilde{c}^2 \tilde{m}_1^2 + \tilde{s}^2 \tilde{m}_2^2 \,, \label{eq:seriesrel1}\\
s^2 m_1^2 + c^2 m_2^2 + A \rho &\simeq \tilde{s}^2 \tilde{m}_1^2 + \tilde{c}^2 \tilde{m}_2^2 \,, \label{eq:seriesrel5}
\end{align}
which, combined, lead to
\begin{equation}
\tilde{m}_1^2 + \tilde{m}_2^2 \simeq m_1^2 + m_2^2 + A + 2 A \rho \,. \label{eq:seriestsum}
\end{equation}
In order to find the two real \emph{approximate} eigenvalues, \emph{i.e.}~the effective neutrino mass-squareds $\tilde{m}_1^2 = \tfrac{1}{2} (\tilde{m}_1^2 + \tilde{m}_2^2 - \Delta \tilde{m}^2)$ and $\tilde{m}_2^2 = \tfrac{1}{2} (\tilde{m}_1^2 + \tilde{m}_2^2 + \Delta \tilde{m}^2)$, we need to compute $\Delta \tilde{m}^2$, which can be achieved by eq.~(\ref{eq:freq}). Series expanding eq.~(\ref{eq:freq}) in the small parameter $\sigma$, we obtain
\begin{align}
\Delta \tilde{m}^2 &= \sqrt{A^2 + (\Delta m^2)^2 - 2 A \Delta m^2 \cos(2\theta)} \nonumber\\
&+ \frac{2 A \Delta m^2 \sin(2\theta)}{\sqrt{A^2 + (\Delta m^2)^2 - 2 A \Delta m^2 \cos(2\theta)}} \sigma + {\cal O}(\sigma^2) \,. \label{eq:seriesfreq}
\end{align}
Observing that $A^2 + (\Delta m^2)^2 - 2 A \Delta m^2 \cos(2\theta) = (\Delta m^2 \sin(2\theta))^2 + (\Delta m^2 \cos(2\theta) - A)^2$ and using eqs.~(\ref{eq:theta'}) and (\ref{eq:m'}), we can therefore write eq.~(\ref{eq:seriesfreq}) as
\begin{equation}
\Delta \tilde{m}^2 = {\Delta m'}^2 + 2 A \sin(2\theta') \sigma + {\cal O}(\sigma^2) \,. \label{eq:seriestfreq}
\end{equation}
Thus, using eqs.~(\ref{eq:seriestsum}) and (\ref{eq:seriestfreq}), we find the two real \emph{app\-roximate} eigenvalues of $H_{\rm eff}$,
\begin{align}
\tilde{m}_1^2 &\simeq \frac{1}{2} (m_1^2 + m_2^2 + A - {\Delta m'}^2) + A \rho - A \sin(2\theta') \sigma \,, \label{eq:seriestm12}\\
\tilde{m}_2^2 &\simeq \frac{1}{2} (m_1^2 + m_2^2 + A + {\Delta m'}^2) + A \rho + A \sin(2\theta') \sigma \,. \label{eq:seriestm22}
\end{align}
Furthermore, series expanding eq.~(\ref{eq:amp}), we obtain the effective mixing angle,
\begin{align}
&\sin^2(2\tilde{\theta}) = \sin^2(2\theta') \nonumber\\
&+ \frac{4A (A - \Delta m^2 \cos(2\theta))^2 \sin(2\theta')}{({\Delta m'}^2)^3} \sigma + {\cal O}(\sigma^2) \,,
\end{align}
which is consistent with eq.~(\ref{eq:ampPT}) if the MSW condition~({\ref{eq:MSW}}) is satisfied. Indeed, eqs.~(\ref{eq:seriestm12}) and (\ref{eq:seriestm22}) are also consistent with eqs.~(\ref{eq:seriesm12}) and (\ref{eq:seriesm22}), since ${\Delta m'}^2 \sin(2\theta') = \Delta m^2 \sin(2\theta)$ always holds. In the vacuum limit, we of course have $\tilde{m}_1^2 = m_1^2$, $\tilde{m}_2^2 = m_2^2$, and $\tilde{\theta} = \theta$. Thus, when $A$ for an experimental setup is not close to the MSW resonance, we cannot use the results in eqs.~(\ref{eq:lambda})--(\ref{eq:freqPT}) and (\ref{eq:seriesm12})--(\ref{eq:singlecond2}). In this case, we need to use perturbation theory in the small parameters $\rho$, $\sigma$, and $\varphi$ in order to find two real \emph{approximate} eigenvalues of $H_{\rm eff}$, which are given by eqs.~(\ref{eq:seriestm12}) and (\ref{eq:seriestm22}).

Naturally, the novelty of using a non-Hermitian PT symmetric effective Hamiltonian can be readily extended to more than two lepton flavors as well as other classes of the effective Hamiltonian itself. It is also beyond the scope of this Letter to investigate the phenomenology of the presented non-Hermitian PT symmetric two-flavor neutrino oscillation formalism, but it would certainly be interesting to perform such investigations for present and future neutrino oscillation experiments.
Furthermore, it is obvious that the new model parameters (introduced by the non-Hermitian PT symmetric Hamiltonian) will mimick (or ``fake'') effects of the old and ordinary fundamental neutrino oscillation parameters \cite{Blennow:2005yk}, and such mimicking effects could also be explored in future works.

In summary, we have shown that a model for two-flavor neutrino oscillations in matter based on a non-Hermitian PT symmetric Hamiltonian described by three real parameters $\rho$, $\sigma$, and $\varphi$ has a real spectrum if either the condition $A^2 [ 4 (\rho \sin \varphi \mp \sigma)^2 + 1] < (\Delta m^2)^2$ is fulfilled or $\rho$, $\sigma$, and $\varphi$ are small. In the first case, the spectrum is exact and the amplitude of the neutrino transition probability is maximal, since the effective leptonic mixing is always maximal due to the fact that the ordinary MSW condition has to be ``automatically'' satisfied, whereas in the second case, the spectrum is approximate. In conclusion, a PT symmetric model opens up the window to ``new physics'', but the additional parameters need to be small in order for the effects to be sub-leading only, since neutrino oscillations are considered to be the most plausible description of neutrino flavor transitions.

\acknowledgments
I am grateful to Avadh Saxena for bringing the seminal work on PT symmetric quantum mechanics by Bender and Boettcher to my attention.

\end{document}